\documentstyle[prl,aps,preprint,epsf]{revtex}

\tighten
\begin{document}
\draft
\title{A theoretical investigation of 
ferromagnetic tunnel junctions with 4-valued conductances
}
\author{
Satoshi~\textsc{Kokado}$^1$ 
\footnote{Electronic mail:
satoshi-kokado@aist.go.jp}
and Kikuo~\textsc{Harigaya}$^{1,2}$
\footnote{Electronic mail:
k.harigaya@aist.go.jp}
}
\address{
$^1$Nanotechnology Research Institute, AIST, Tsukuba 305-8568, Japan \\
$^2$Synthetic Nano-Function Materials Project, AIST, Tsukuba 305-8568, Japan
}
\date{\today}
\maketitle

\begin{abstract}
In considering a novel function 
in ferromagnetic tunnel junctions consisting of 
ferromagnet(FM)/barrier/FM junctions, 
we theoretically investigate 
multiple valued (or multi-level) cell property, 
which is in principle 
realized by sensing conductances of four states 
recorded with magnetization configurations of two FMs; 
that is, (up,up), (up,down), (down,up), (down,down). 
To obtain such 4-valued conductances, 
we propose FM1/spin-polarized barrier/FM2 junctions, 
where the FM1 and FM2 are different ferromagnets, 
and the barrier has spin dependence. 
The proposed idea is applied to  
the case of the barrier having localized spins. 
Assuming that all the localized spins are pinned 
parallel to magnetization axes of the FM1 and FM2, 
4-valued conductances are explicitly obtained 
for the case of many localized spins. 
Furthermore, objectives for an ideal spin-polarized barrier 
are discussed. 
\end{abstract}


\section{Introduction}
Ferromagnetic tunnel junctions (FTJ) 
such as ferromagnet(FM)/barrier/FM junctions~\cite{TMR,60,semi,delta} 
have been recently applied to 
elements in magnetic random access memories (MRAM) 
because of their tunnel magnetoresistance (TMR) effect, 
which appears when an applied magnetic field changes an angle 
between magnetizations of two FMs. 
In the practical use, a spin-valve type~\cite{spin-valve} is usually adopted. 
For the writing process, 
the magnetization of only one side of the FM is changed 
under the applied field, 
so the magnetization configurations 
between two FMs are parallel (P) or anti-parallel (AP); 
for the reading process, 
the difference in resistance between the P and AP cases is used.

For the FTJ, 
much effort has been made to mainly increase TMR ratio, 
which is defined by 
$(\Gamma_P - \Gamma_{AP})/\Gamma_{AP}$, where $\Gamma_{P(AP)}$ is 
conductance of the P (AP) case. 
Experimentally, 
the TMR ratio of Co-Fe/Al-O/Co-Fe junctions reached 60\% 
at room temperature~\cite{60} 
in investigations of the influence of the fabrication method for oxide barrier 
on the TMR effect, 
epitaxially grown Ga$_{1-x}$Mn$_x$As/AlAs/Ga$_{1-x}$Mn$_x$As junctions 
have exhibited the TMR ratio more than 70\% at 8 K~\cite{semi}, 
and the TMR ratio of Co/Fe-doped Al$_2$O$_3$/Ni$_{80}$Fe$_{20}$ junctions 
has been successfully enhanced as compared to 
those with undoped Al$_2$O$_3$~\cite{delta}. 
Theoretically, 
the TMR ratio for epitaxial Fe/MgO/Fe(001) junctions 
became more than 300\% 
because of coherent tunneling 
with conservation of intralayer momentum~\cite{CC}, 
and 
TMR ratios for junctions having 
a sheet with 100\% spin polarized dopants in the barrier~\cite{del_th1} 
and the Fe-doped barrier~\cite{del_th} 
were enhanced because of 
spin polarization of resonant states inside the barrier and 
spin dependent energy levels inside it, respectively. 
Furthermore, double barrier junctions of 
FM/barrier/nonmagnetic layer/barrier/FM 
reached a TMR ratio of more than 5000\% at a specific thickness 
of the nonmagnetic layer, 
as a result of spin dependent conduction 
through resonant levels in the layer~\cite{Wil1}.

By comparing such FTJ with elements~\cite{set1} 
in other memories such as Flash memory~\cite{MLC}, 
however, we find out that there are few studies on 
multiple valued (or multi-level) cell property~\cite{MLC}, which 
allows many bits to be stored in each memory cell, 
and reduces memory cell size by 1/(the number of bits). 
If such a property is included in the FTJ, 
they will function as a more efficient memory cell than 
the conventional one.

In this paper, 
we theoretically investigate 
the multiple valued cell property~\cite{MLC} in the FTJ. 
In a possible scheme, 
recorded states are supposed to be 
four magnetization configurations of two FMs consisting of 
(up,up), (up,down), (down,up), (down,down), 
which are obtained by applying magnetic fields to respective FMs. 
The FTJ correspond to 2bits memory cells. 
Then, in order to sense all the states, 
4-valued conductances corresponding to the respective states are 
obviously necessary. 
By paying attention to the magnitude of total magnetization 
in the whole system, 
a model to obtain such conductances is considered to be 
FM1/spin-polarized barrier (SPB)/FM2 junctions, 
where the FM1 and FM2 have different spin polarizations 
and magnetization in the barrier is pinned, 
according to the following procedure.  
First, a difference of conductances 
between (up,up) and (down,down) will appear, 
if the barrier is a SPB. 
Second, a difference between (up,down) and (down,up) 
will be obtained by introducing the FM1 and FM2, 
in addition to the SPB. 
In actual calculations, the junctions 
exhibit 4-valued conductances. 
Further, 
the proposed idea is applied to 
the case of the barrier having localized spins. 
When all the localized spins are pinned 
parallel to the magnetization axes of FM1 and FM2, 
4-valued conductances are explicitly obtained 
in the case of many localized spins. 
Finally, objectives for an ideal SPB 
to certainly observe such conductances 
are discussed.

\section{3-valued and 4-valued conductances}
We first investigate the conductance of the FM1/SPB/FM2 junctions, 
where the barrier merely has spin dependent height 
due to the pinned magnetization, 
and its material is not specified for general discussions. 
To simply find intrinsic properties, 
the one-dimensional systems are adopted. 
Within the Green's function technique~\cite{theory,kokado}, 
the conductance is given by
\begin{eqnarray}
\label{cond}
&&\Gamma = 
\frac{4 \pi^2 e^2 }{h} \sum_{\sigma=\uparrow,\downarrow}
\sum_{\sigma'=\uparrow,\downarrow}
T_{\sigma,\sigma'}
D_{1,\sigma}(E_{\mbox{\tiny F}}) 
D_{2,\sigma'}(E_{\mbox{\tiny F}}), 
\end{eqnarray}
where 
$\sigma$ is spin of a tunnel electron, and 
$D_{1(2),\sigma}(E)$ denotes the local density-of-states (DOS) 
at an interfacial layer in FM1(2) 
at the Fermi level, $E_{\mbox{\tiny F}}$. 
$T_{\sigma,\sigma'}$ 
is a spin dependent transmission coefficient 
including the spin-flip process of $\sigma \ne \sigma'$, 
and is proportional to $|G_{\sigma,\sigma'}|^2$, 
where $G_{\sigma,\sigma'}$ is 
the $(\sigma,\sigma')$ component of 
an element between both edge sites in the barrier 
for the Green's function of the whole system~\cite{theory,kokado}. 
As the barrier height becomes higher, 
$T_{\sigma,\sigma'}$ decreases.

Using eq. (\ref{cond}), 
we obtain conductances for respective configurations, 
where the magnetization state of FM1 (FM2) 
is represented by $m1$ ($m2$), which is $\Uparrow$ or $\Downarrow$. 
By introducing $\gamma_{m1,m2} = \Gamma_{m1,m2}/\Gamma_{\Uparrow,\Uparrow}$, 
we have 
\begin{eqnarray}
&&\gamma_{\Uparrow,\Uparrow}= 
\frac{\Gamma_{\Uparrow,\Uparrow}}{\Gamma_{\Uparrow,\Uparrow}}=1, \\
&&\gamma_{\Downarrow,\Downarrow}= 
\frac{\Gamma_{\Downarrow,\Downarrow}}{\Gamma_{\Uparrow,\Uparrow}}
=\displaystyle{
\frac{d_1 d_2+t+t'd_2 +t''d_1}{1+t d_1 d_2+t'd_2+t''d_1}},  \\
&&\gamma_{\Uparrow,\Downarrow}= 
\frac{\Gamma_{\Uparrow,\Downarrow}}{\Gamma_{\Uparrow,\Uparrow}}
=\displaystyle{
\frac{d_2+td_1+t'+t''d_1 d_2}{1+t d_1 d_2+t'd_2+t''d_1}}, \\
&&\gamma_{\Downarrow,\Uparrow}= 
\frac{\Gamma_{\Downarrow,\Uparrow}}{\Gamma_{\Uparrow,\Uparrow}}
=\displaystyle{
\frac{d_1 +t d_2 +t'd_1 d_2+t''}{1+td_1 d_2+t'd_2+t''d_1}}, 
\end{eqnarray}
with 
$t=T_{\downarrow,\downarrow}/T_{\uparrow,\uparrow}$, 
$t'=T_{\uparrow,\downarrow}/T_{\uparrow,\uparrow}$, 
$t''=T_{\downarrow,\uparrow}/T_{\uparrow,\uparrow}$, 
$d_1=D_{1,m}(E_{\mbox{\tiny F}})/D_{1,M}(E_{\mbox{\tiny F}})$, 
and $d_2=D_{2,m}(E_{\mbox{\tiny F}})/D_{2,M}(E_{\mbox{\tiny F}})$, 
where 
$D_{1(2),M}(E_{\mbox{\tiny F}})$ 
and $D_{1(2),m}(E_{\mbox{\tiny F}})$ correspond to 
local DOSs at $E_{\mbox{\tiny F}}$ 
for majority spin and for minority spin 
of the case of $\Uparrow,\Uparrow$, respectively. 
In addition, the relation of $0 \le d_1, d_2 \le 1$ is satisfied. 
Note here that 
the case of $t$=1 and $t'=t''=0$ has 2-valued conductances, 
and then 
$1-\gamma_{\Uparrow,\Downarrow}$ (or $1-\gamma_{\Downarrow,\Uparrow}$) 
results in the Julliere model~\cite{Julliere}.

By putting $t'=t''=0$, we investigate $1-t$ 
$[=(T_{\uparrow,\uparrow}-T_{\downarrow,\downarrow})/T_{\uparrow,\uparrow}]$ 
dependence of $\gamma_{m1,m2}$. 
In particular, we study the influence on $\gamma_{m1,m2}$ of a change 
from the non-spin-polarized barrier to the spin-polarized one.

The FM/SPB/FM junctions with 
$d_1=d_2$ and $t \ne d_1$ exhibit 3-valued conductances, 
while those with $t=d_1$ have 2-valued conductances. 
Figure 1(a) shows $\gamma_{m1,m2}$ with $d_1=d_2=0.41$~\cite{sp}, 
which are determined by the least squares method 
such that ($\gamma_{\Uparrow,\Downarrow}$,$\gamma_{\Downarrow,\Downarrow}$) 
becomes (0.55,0.1) at $1-t$=1. 
The quantities 
$\gamma_{\Downarrow,\Downarrow}$ and $\gamma_{\Uparrow,\Downarrow}$ 
decrease nearly linearly with increasing $1-t$, 
while $\gamma_{\Downarrow,\Downarrow}$ crosses $\gamma_{\Uparrow,\Downarrow}$ 
at $1-t=1-d_1$, 
and $\gamma_{\Downarrow,\Downarrow}$ is smaller than 
$\gamma_{\Uparrow,\Downarrow}$ for $1-t > 1-d_1$. 
The differences of $\gamma_{m1,m2}$ 
between all the configurations become large near $1-t$=1.

The FM1/SPB/FM2 junctions 
with $d_1 < d_2$ and $t \ne d_1,d_2$ exhibit 4-valued conductances, 
while those with $t=d_1$ or $d_2$ have 3-valued conductances. 
Figure 1(b) shows $\gamma_{m1,m2}$ with $d_1$=0.38 and $d_2$=0.58~\cite{sp}, 
which is determined by the least squares method 
such that 
($\gamma_{\Uparrow,\Downarrow}$,$\gamma_{\Downarrow,\Uparrow}$,$\gamma_{\Downarrow,\Downarrow}$) becomes (0.7,0.4,0.1) at $1-t$=1. 
The quantities 
$\gamma_{\Downarrow,\Downarrow}$, $\gamma_{\Uparrow,\Downarrow}$, 
and $\gamma_{\Downarrow,\Uparrow}$ decrease with $1-t$, 
while $\gamma_{\Downarrow,\Downarrow}$ crosses 
$\gamma_{\Uparrow,\Downarrow}$ at $1-t=1-d_2$ and 
$\gamma_{\Downarrow,\Uparrow}$ at $1-t=1-d_1$. 
The differences between all $\gamma_{m1,m2}$ 
are found obviously near $1-t$=1.

\section{Application to the barrier having localized spins}
As an example to obtain $T_{\sigma,\sigma'}$ difinitively, 
we focus on the barrier having localized spins 
due to magnetic ions or ions in magnetic particles, 
which are configured linearly and with same interval [see Fig. 2(a)]. 
We investigate $\gamma_{m1,m2}$ 
through a quantum well potential structure with dependence on spin 
by assuming 
that the potential of conduction level at the magnetic ion site 
is lower than that in the original barrier part, 
and magnetic couplings 
between the localized spins and the magnetizations in the FMs are so small 
that influence on the spin dependent conduction might be negligible. 
We use a single orbital tight-binding model 
with nearest neighbor transfer integrals, 
and take into account exchange interactions 
between the tunnel electron spin and localized spins~\cite{del_th}, 
while couplings between localized spins are not specified 
because their ferromagnetic, antiferromagnetic, and canted states 
are considered. 
In the unit of magnitudes of the transfer integral, 
the Hamiltonian in the barrier is given by,
\begin{eqnarray}
\label{ham}
&&{\cal H} =
   e \sum_{i,\sigma} 
   | i,\sigma \rangle \langle i,\sigma |
   -\sum_{i,\sigma}
   (|i,\sigma \rangle \langle i+1,\sigma|+{\rm h.c.}) \nonumber \\
&&\hspace*{0.7cm}-J \sum_{i,\sigma,\sigma'} 
{\bf \sigma}_{\sigma,\sigma'} \cdot {\bf S}_i
| i,\sigma \rangle \langle i,\sigma' |,
\end{eqnarray}
where $|i,\sigma \rangle$ is an orbital with spin-$\sigma$ 
($=\uparrow {\rm or} \downarrow$) at $i$-th site, 
and $e$ denotes the on-site energy. 
Further, $J$ is a ferromagnetic exchange integral 
with positive sign~\cite{del_th}, 
${\bf \sigma}_{\sigma,\sigma'}$ is 
the $(\sigma,\sigma')$ component of the Pauli matrix, and 
${\bf S}_i$ $[=(S_{i,x},S_{i,y},S_{i,z})]$ is regarded as 
a classical spin at $i$-th site.

When the number of localized spins is $n$ ($>1$), 
$T_{\sigma,\sigma'}$ is proportional to 
$|\langle 1,\sigma |G | n,\sigma' \rangle |^2$, 
where $G$ is treated approximately as a bare Green's function 
only in the barrier 
on the assumption that the self energy correction is negligibly small 
reflecting very small couplings between the FMs and the SPB. 
Using eq. (\ref{ham}), 
$G$ is written as 
\begin{eqnarray}
\label{Green1}
&&G = 
({\cal G}^{-1} - {\cal T} )^{-1}, 
\end{eqnarray}
with ${\cal T}=-\sum_{i,\sigma}
   ( |i,\sigma \rangle \langle i+1,\sigma|+{\rm h.c.})$, and 
\begin{eqnarray}
&&{\cal G} =\left[ (E_{\mbox{\tiny F}}  - e + {\rm i}0^+) 
\sum_{i,\sigma} | i,\sigma \rangle \langle i,\sigma | \right. \nonumber \\ 
&& \hspace*{1.5cm}\left. + J \sum_{i,\sigma,\sigma'} 
{\bf \sigma}_{\sigma,\sigma'} \cdot {\bf S}_i
| i,\sigma \rangle \langle i,\sigma' | \right]^{-1}. 
\end{eqnarray}
Here, ${\cal G}$ is 
a Green's function on sites unconnected to the other sites, and 
is exactly expressed as 
\begin{eqnarray}
{\cal G} = \sum_{i,\sigma,\sigma'} 
{\cal G}_i^{(\sigma,\sigma')} | i,\sigma \rangle \langle i,\sigma' |, 
\end{eqnarray}
with 
\begin{eqnarray}
&&{\cal G}_i^{(\sigma,\sigma')} = \frac{g^{-1} \delta_{\sigma,\sigma'} - J 
{\bf \sigma}_{\sigma,\sigma'}  \cdot {\bf S}_i }{g^{-2} - J^2 S^2} \nonumber \\
&& \hspace*{1.1cm}
=\frac{1}{2} \left( \delta_{\sigma,\sigma'} + \displaystyle{\frac{{\bf \sigma}_{\sigma,\sigma'}\cdot {\bf S}_i }{S}} \right) \frac{1}{g^{-1}+J S} \nonumber \\
&&\hspace*{1.5cm}+ \frac{1}{2} \left( \delta_{\sigma,\sigma'} -  \displaystyle{\frac{{\bf \sigma}_{\sigma,\sigma'}\cdot {\bf S}_i }{S}} \right) \frac{1}{g^{-1}-J S}, 
\end{eqnarray}
$g={(E_{\mbox{\tiny F}} -e + {\rm i} 0^+)}^{-1}$, and ${\bf S}_i^2=S^2$. 
The first and second terms correspond to components of 
a lower level with $e-JS$ 
and of an upper level 
with $e+JS$, respectively. 
For $J$=0, ${\cal G}_i^{(\sigma,\sigma')}$ 
certainly has the non-spin-polarized feature. 
For $J \ne 0$, when ${\bf S}_i$ is along the quantum axis, i.e., 
${\bf \sigma}_i \cdot {\bf S}_i/S=\sigma_{i,z}$, 
${\cal G}_i^{(\sigma,\sigma')}$ has only the spin-up (-down) component 
for the lower (upper) level, 
which represents the largely spin-polarized feature, 
while when ${\bf S}_i$ perpendicular to the quantum axis, 
${\cal G}_i^{(\sigma,\sigma')}$ shows no difference 
between the spin-up and spin-down in both the levels, 
which represents the non-spin-polarized feature.

In this calculation, 
we utilize the same set ($d_1,d_2$)=(0.38,0.58) as above, 
and choose $e-E_{\mbox{\tiny F}}$=3.5, where 
a condition of $e-JS > E_{\mbox{\tiny F}}$ is used, 
based on strong on-site Coulomb repulsions 
at magnetic ion sites~\cite{del_th}. 
Further, localized spins are assumed to exist parallel to $yz$-plane, 
and an angle at odd (even) sites between localized spins and $z$-axis 
is written as $\theta_o$ ($\theta_e$) [see Fig. 2(b)].

In Fig. 3(a), 
we show the $JS$ dependence of $\gamma_{m1,m2}$ 
for $\theta_o/\pi=\theta_e/\pi=0$. 
At $JS$=0, 
the difference of $\gamma_{m1,m2}$ exists only between the P and AP cases. 
With increasing $JS$, 
each $\gamma_{m1,m2}$ approaches to its saturation value, 
which corresponds just to each $\gamma_{m1,m2}$ at $1-t=1$ shown in Fig. 1(b). 
It is worth noting that 
each $\gamma_{m1,m2}$ of $n$=8 saturates more rapidly than that of $n$=2. 
Such behavior reflects the fact that $t$ of $n$=8 
decreases with $JS$ more drastically than 
that of $n$=2 
according to 
\begin{eqnarray}
\label{t}
t = 
\frac{\sinh^2 \kappa_\downarrow/ \sinh^2 [\kappa_\downarrow (n+1)]}{
\sinh^2 \kappa_\uparrow/ \sinh^2 [\kappa_\uparrow (n+1)]} , 
\end{eqnarray}
with 
$2\cosh \kappa_{\uparrow (\downarrow)} 
= e - (+) JS -E_{\mbox{\tiny F}}$~\cite{t,kokado}. 
When the decay exponentials in the hyperbolic sine of eq. (\ref{t}) 
can be neglected owing to large $\kappa_{\uparrow (\downarrow)}$, 
we have $t \approx {\rm e}^{-2\Delta \kappa n}$, 
where $\Delta \kappa$, defined by $\kappa_\downarrow - \kappa_\uparrow$, 
increases with $JS$~\cite{mathon}.

Figure 3(b) shows the $\theta_o$ (=$\theta_e$) dependence of $\gamma_{m1,m2}$ 
for $JS$=1. 
A condition of $\theta_o/\pi$=$-$0.5 (0) represents localized spins 
oriented in $-y$ ($z$) direction. 
At $\theta_o/\pi=-0.5$, 
only the difference between the P and AP cases is present. 
The difference at $\theta_o/\pi=-0.5$ of $n$=8 
is much smaller than that of $n$=2, 
because the spin-flip tunneling process increases 
owing to an increase of transverse components of localized spins. 
As $\theta_o$ approaches to 0, 
differences of $\gamma_{m1,m2}$ among all the configurations 
become large.

In Fig. 3(c), 
we investigate the $\theta_o$ dependence of $\gamma_{m1,m2}$ 
for $JS$=1 and $\theta_e/\pi=0$. 
The change of $\theta_o$ from $-\pi$ to 0 corresponds to 
that from antiferromagnetic state to ferromagnetic one 
via canted ones. 
The quantity $\gamma_{m1,m2}$ 
exhibits a difference only between the P and AP cases 
at $\theta_o/\pi=-1$; 
they individually behave for $-1 < \theta_o/\pi < 0$; 
and there are large differences between all the configurations 
at $\theta_o/\pi=0$.

\section{Discussions}
We now comment on a comparison with previous works. 
As far as we are aware, resistance versus magnetic fields observed for 
Co/Fe-doped Al$_2$O$_3$/Ni$_{80}$Fe$_{20}$ junctions~\cite{delta}, 
which may be similar to FM1/SPB/FM2 junctions with 
the above mentioned barrier having localized spins, 
have not confirmed 3-valued or 4-valued conductances. 
Note that localized spins of Fe were not pinned, 
and $1-t$ might possibly be small. 
On the other hand, the observed enhancement of several per cent 
in the TMR ratio 
appears to originate from the behavior of $\gamma_{\Uparrow,\Downarrow}$ 
or $\gamma_{\Downarrow,\Uparrow}$ for $1-t \ne 0$, as shown in Fig. 1. 
Also, the increase of the TMR ratio with $JS$ shown in Fig. 3(a) 
agrees qualitatively with 
that of the previous theoretical study~\cite{del_th}.

We here consider an ideal SPB 
to certainly observe 3-valued or 4-valued conductances. 
For the barrier having magnetic particles, 
we give two objectives, 
which are 
\begin{itemize}
\item[(i)] to strongly pin the magnetization 
of magnetic particles in the barrier 
parallel to magnetization axes of FMs, 
and 
\item[(ii)] to diminish the magnetic couplings 
between magnetic particles and FMs. 
\end{itemize}
As for (i), we propose magnetic particles 
having a coercive field higher than those of FMs. 
Also, it may be effective to apply 
exchange bias from an antiferromagnet~\cite{spin-valve} 
to the magnetic particles. 
Here, the antiferromagnet is 
located not between the FMs and the SPB 
but at the side of the SPB in the FTJ, 
meaning that few conduction electrons may flow in the antiferromagnet 
because it is not connected to FM electrodes. 
For (ii), 
we suppose that 
distances between magnetic particles and FMs 
should be well controlled 
so that 
the magnetic dipole-dipole interactions between them 
become very small and 
have little influence on the spin dependent conduction.

As a realistic SPB, which could satisfy such objectives, 
we bear in mind of 
a carbon nanotube encapsulating magnetic particles~\cite{nt,presen} 
by the following reasons: 
Size of magnetic particles may be close to 
that of a single domain particle 
by tuning conditions of fabrication. 
Note that the coercive field was recently observed as about 0.5 kOe at 300 K 
even for Fe particles encapsulated 
with non-single domain size of about 70 nm~\cite{nt}. 
Further, magnetic particles can be encapsulated 
not only at edges of the nanotube~\cite{nt} 
but also in the inner region~\cite{nt1}, 
where 
distances between the particles and the edges 
may be tunable 
by controlling nanotube growth processes. 
It is also mentioned that 
the nanotube itself has very long spin-flip scattering lengths 
which extend 130 nm at least~\cite{Tsuka}.

As another SPB, 
we propose a ferromagnetic barrier such as EuS, 
where bulk EuS has a band gap of 1.65 eV and 
the exchange splitting of conduction band of 0.36 eV~\cite{Hao}. 
In fact, it has been shown experimentally that 
the EuS tunnel barrier can be used as 
a highly efficient spin filter~\cite{LeC}, 
and it has been theoretically found that 
the ferromagnetic barrier 
largely contributes to an increase of the TMR ratio 
in double ferromagnetic barrier junctions, where 
external and central electrodes are nonmagnetic~\cite{Wil1,Filip,Wil2}. 
When such a ferromagnetic barrier is used as the SPB, 
we design FM1/barrier/SPB/barrier/FM2 junctions, where 
the antiferrmagnet may be layered at the side of the SPB 
in order to pin the magnetization of the SPB strongly.

>From the viewpoint of device applications, 
we anticipate that the 2bits/cell MRAM will be realized 
using 4-valued conductances. 
At the same time, we mention that 
the FTJ with magnetization reversals 
between $\Uparrow,\Uparrow$ and $\Downarrow,\Downarrow$
show larger TMR ratio compared with 
the conventional spin-valve type~\cite{spin-valve}, 
in the case of a largely spin-polarized barrier.

\section{Conclusion}
We have theoretically investigated the multiple valued cell property, 
which is in principle realized by sensing three or four states 
recorded with magnetization configurations of two FMs. 
The FM/SPB/FM junctions have 
3-valued conductances to sense three states, 
and the FM1/SPB/FM2 junctions, 
where the FM1 and FM2 have the different local DOSs at $E_{\mbox{\tiny F}}$, 
have 4-valued conductances to sense four states. 
When the barrier has localized spins, 
differences among those conductances are strongly influenced by 
the magnitude of the interaction 
between the tunnel electron spin and localized spins, 
and the directions of the localized spins. 
In the case of many localized spins, 
respective conductances rapidly approach saturation values. 
Further, 
objectives for the ideal SPB 
to observe such conductances 
have been given and considered. 
We expect that 
the present proposal on 4-valued conductances 
will contribute a great deal to the realization of the 2bits/cell MRAM 
in the near future.

\section{Acknowledgements}
This work has been supported by 
Special Coordination Funds for Promoting Science and Technology, Japan. 
One of the authors (K.H.) acknowledges 
the partial financial support from NEDO via 
Synthetic Nano-Function Materials Project, AIST, Japan, too.

\begin{figure}
\caption{
$\gamma_{m1,m2}$ versus $1-t~[=(T_{\Uparrow,\Uparrow} - T_{\Downarrow,\Downarrow})/T_{\Uparrow,\Uparrow}]$ 
for $t'=t''=0$. 
(a) The FM/spin-polarized barrier (SPB)/FM junctions 
with 3-valued conductances. ($d_1,d_2$)= (0.41,0.41). 
(b) The FM1/SPB/FM2 junctions with 4-valued conductances, 
where the FM1 and FM2 have different local DOSs at $E_{\mbox{\tiny F}}$. 
($d_1,d_2$)=(0.38,0.58). 
Here, 
sets of two arrows represent magnetization configurations of two FMs. 
}
\end{figure} 
\begin{figure}
\caption{
(a) A schematic illustration of the FM1/SPB/FM2 junctions, 
where localized spins are linearly configured in the barrier. 
Open (solid) arrays represent the 
FM's magnetizations (localized spins). 
(b) 
Configurations of the localized spins projected in $yz$-plane. 
$\theta_o$ ($\theta_e$) 
is the angle at odd (even) sites 
between the localized spins and the $z$-axis. 
}
\end{figure} 
\begin{figure}
\caption{
$\gamma_{m1,m2}$ of the FTJ in Fig. 2. 
($d_1,d_2$)= (0.38,0.58). 
The notation follows that in Fig. 1. 
(a) $JS$ dependence up to $JS$=1.49 for $\theta_o$=$\theta_e$=0. 
(b) $\theta_o$ ($=\theta_e$) dependence for $JS$=1. 
(c) $\theta_o$ dependence for $JS$=1 and $\theta_e$=0. 
Here, the unit of $J$ is the magnitude of the transfer integral. 
}
\end{figure} 


\begin{thebibliography}{99} %
\bibitem{TMR}
T. Miyazaki and N. Tezuka, 
J. Magn. Magn. Mater. {\bf 139} (1995) L231; 
J. S. Moodera, L. R. Kinder, T. M. Wong, and R. Meservey, 
Phys. Rev. Lett. {\bf 74} (1995) 3273. 
\bibitem{60}
M. Tsunoda, K. Nishikawa, S. Ogata, and M. Takahashi, 
Appl. Phys. Lett. {\bf 80} (2002) 3135. 
\bibitem{semi}
M. Tanaka and Y. Higo, 
Phys. Rev. Lett. {\bf 87} (2001) 026602. 
\bibitem{delta}
R. Jansen and J. S. Moodera, 
Appl. Phys. Lett. {\bf 75} (1999) 400. 
\bibitem{spin-valve}
S. S. Parkin, K. P. Roche, M. G. Samant, P. M. Rice, R. B. Beyers, 
R. E. Scheuerlein, E. J. O'Sullivan, S. L. Brown, 
J. Bucchigano, D. W. Abraham, Yu. Lu, M. Rooks, P. L. Trouilloud, 
R. A. Wanner, W. J. Gallagher, J. Appl. Phys. {\bf 85} (1999) 5828. 
\bibitem{CC}
J. Mathon and A. Umerski, Phys. Rev. B {\bf 63}, 220403(R) (2001). 
\bibitem{del_th1}
R. Jansen and J. C. Lodder, Phys. Rev. B {\bf 61} (2000) 5860. 
\bibitem{del_th}
J. Inoue, N. Nishimura, and H. Itoh, 
Phys. Rev. B {\bf 65} (2002) 104433. 
\bibitem{Wil1}
M. Wilczynski and J. Barnas, 
Sens. and Actuators A {\bf 91} (2001) 188. 
\bibitem{set1}
For example, see 
N. Takahashi, H. Ishikuro, and T. Hiramoto, 
Appl. Phys. Lett. {\bf 76} (2000) 209. 
\bibitem{MLC}
R. V. Giridhar, Jpn. J. Appl. Phys. {\bf 35} (1996) 6347. 
\bibitem{theory} 
T. N. Todorov, G. A. D. Briggs, and A. P. Sutton, J. Phys.: Condens. Matter {\bf 5} (1993) 2389. 
\bibitem{kokado}
S. Kokado, M. Ichimura, T. Onogi, A. Sakuma, R. Arai, J. Hayakawa, K. Ito, and Y. Suzuki, Appl. Phys. Lett {\bf 79} (2001) 3986. 
\bibitem{Julliere} 
M. Julliere, 
Phys. Lett. A {\bf 54} (1975) 225. 
\bibitem{sp}
Materials with $d_1$=$d_2$=0.41 and $d_1$=0.38 
correspond nearly to Co, Fe, and Ni$_{80}$Fe$_{20}$, while $d_2$=0.58 is Ni.~
J. S. Moodera and G. Mathon, J. Magn. Magn. Mater. {\bf 200} (1999) 248. 
\bibitem{t} 
A. P. Sutton, M. W. Finnist, D. G. Pettifor, and Y. Ohta, 
J. Phys. C, Solid State Phys. {\bf 21} (1988) 35. 
\bibitem{mathon}
For example, see 
J. Mathon, Phys. Rev. B {\bf 56} (1997) 11810. 
\bibitem{nt}
X. X. Zhang, G. H. Wen, S. Huang, L. Dai, R. Gao, Z. L. Wang, 
J. Magn. Magn. Mater. {\bf 231} (2001) L9. 
\bibitem{presen}
S. Kokado and K. Harigaya, to be published in Physica E.  
\bibitem{nt1}
X. Zhao, S. Inoue, M. Jinno, T. Suzuki, Y. Ando, 
Chem. Phys. Lett. {\bf 373} (2003) 266. 
\bibitem{Tsuka}
K. Tsukagoshi, B. W. Alphenaar, and H. Ago, Nature {\bf 401} (1999) 572. 
\bibitem{Hao}
X. Hao, J. S. Moodera, and R. Meservey, 
Phys. Rev. B {\bf 42} (1990) 8235. 
\bibitem{LeC}
P. LeClair, J. K. Ha, H. J. M. Swagten, J. T. Kohlhepp, C. H. van de Vin, 
and W. J. M. de Jonge, 
Appl. Phys. Lett. {\bf 80} (2002) 625. 
\bibitem{Filip}
A. T. Filip, P. LeClair, C. J. P. Smits, J. T. Kohlhepp, H. J. M. Swagten, 
B. Koopmans, and W. J. M. de Jonge, 
Appl. Phys. Lett. {\bf 81} (2002) 1815. 
\bibitem{Wil2}
M. Wilczy$\acute{\rm n}$ski, J. Barna$\acute{\rm s}$, 
and R. $\acute{\rm S}$wirkowicz, 
phys. stat. sol. (a) {\bf 196} (2003) 109.
\end{thebibliography}
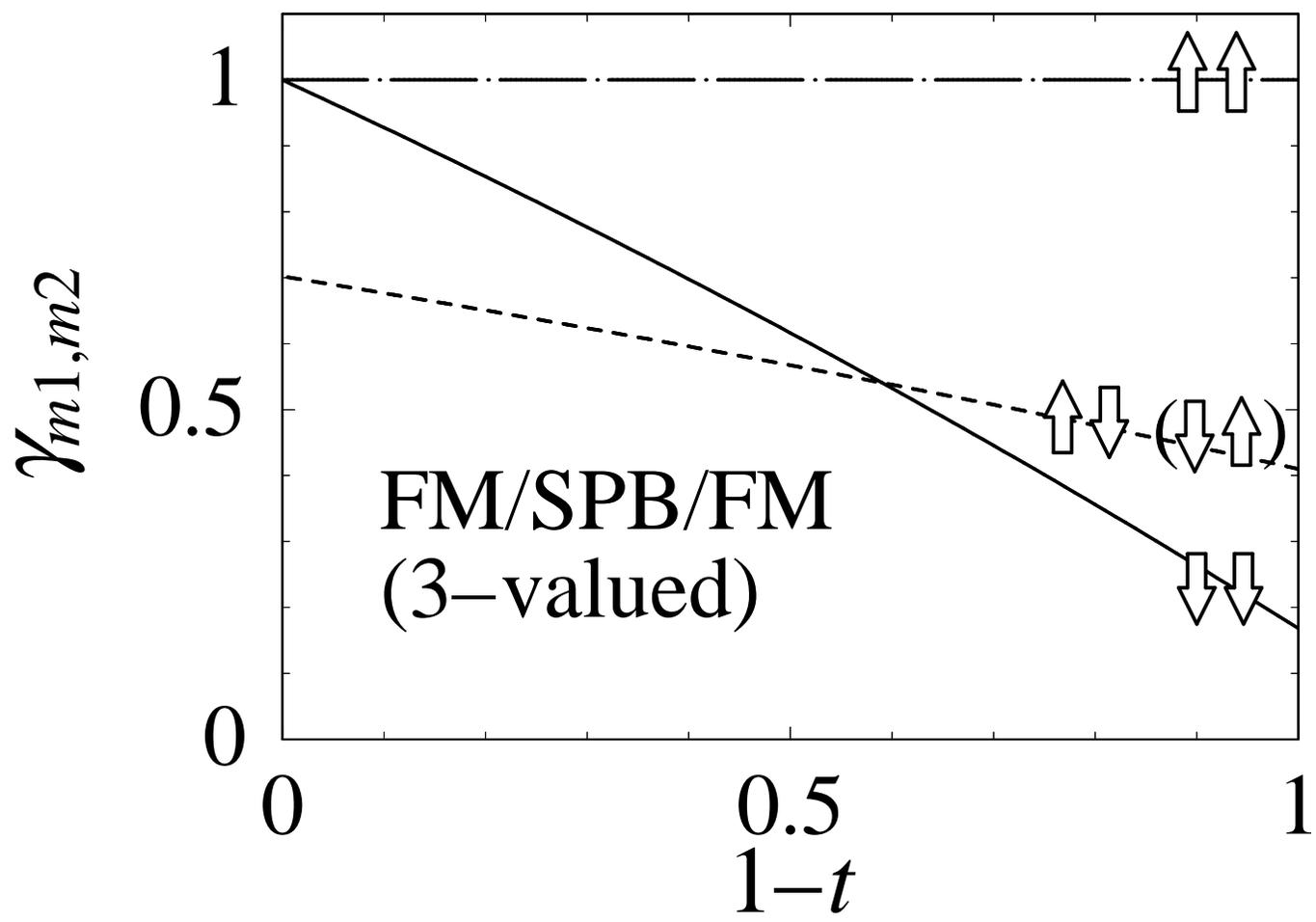
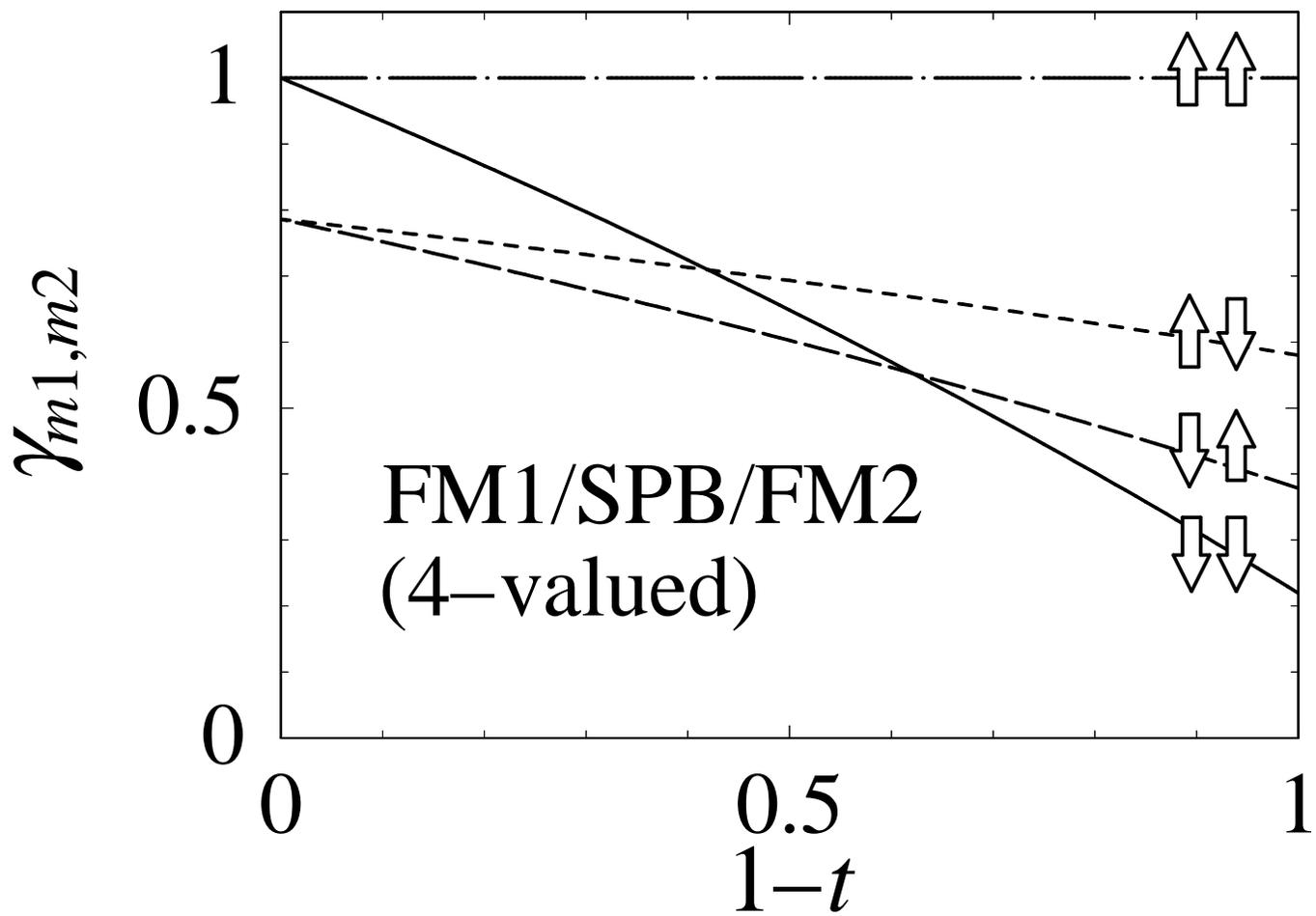
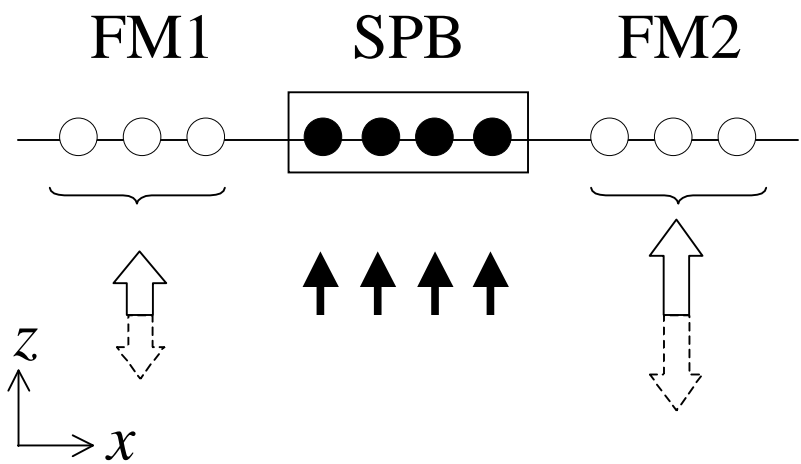
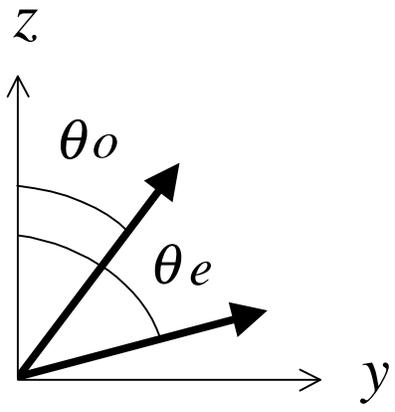
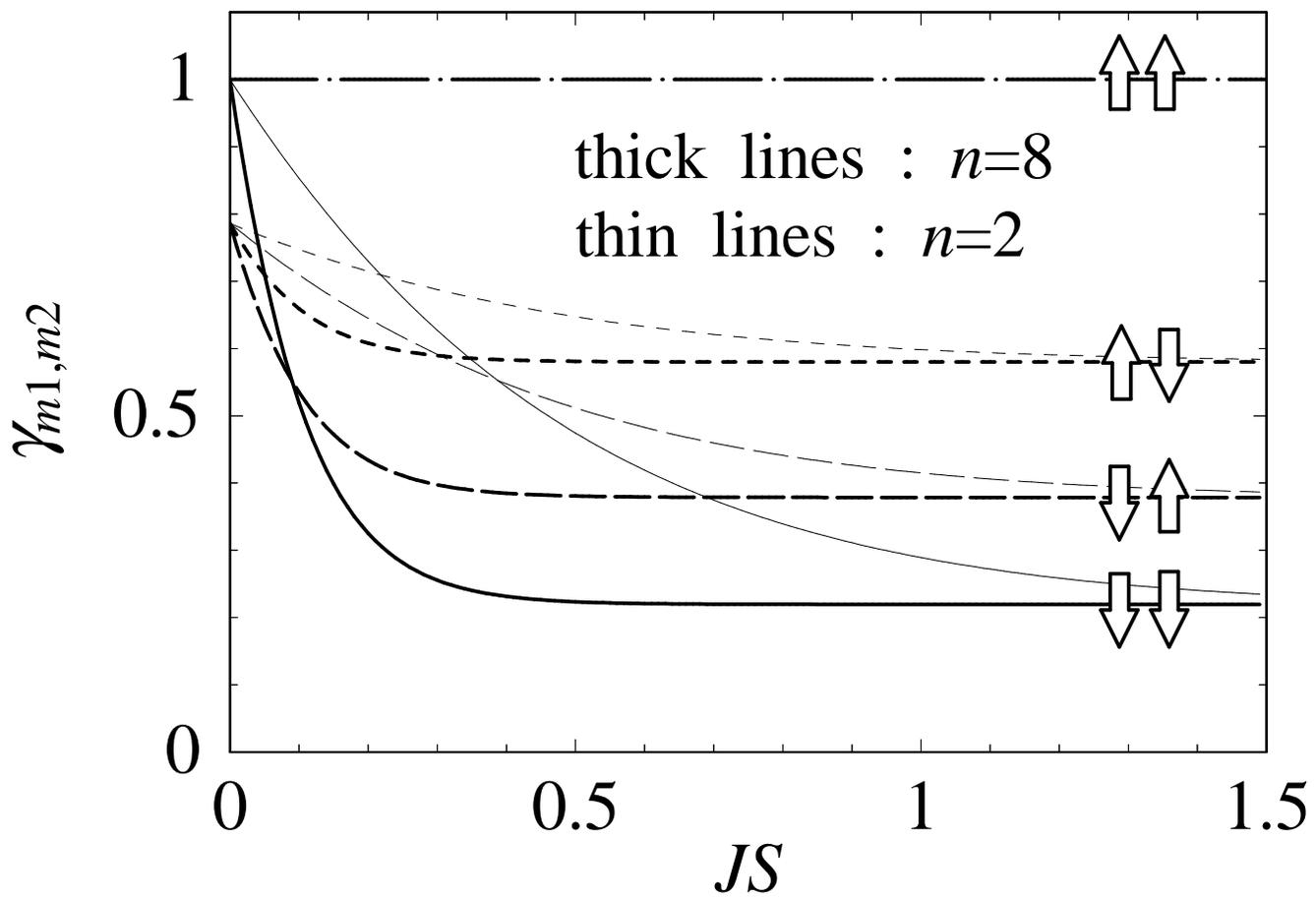
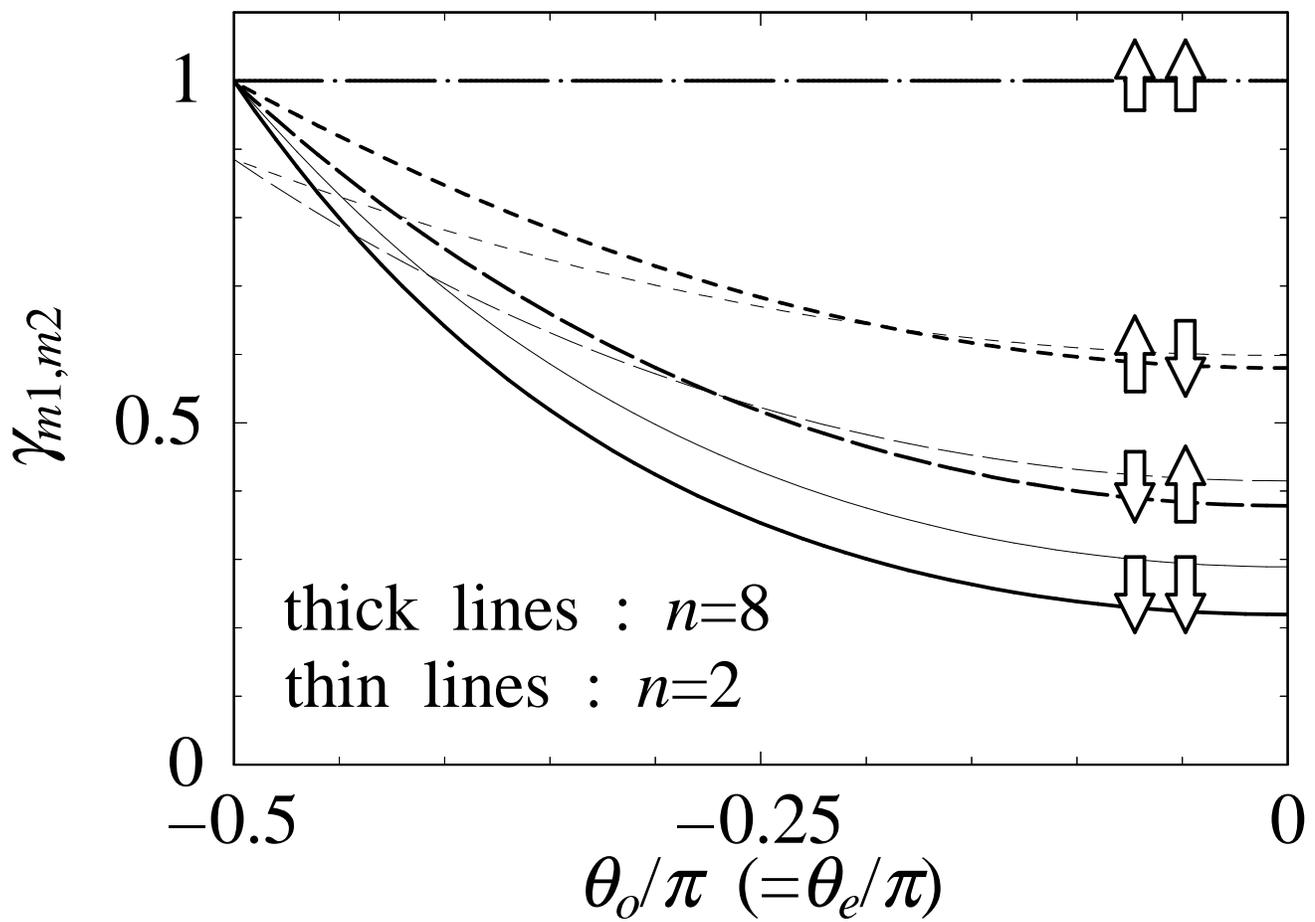
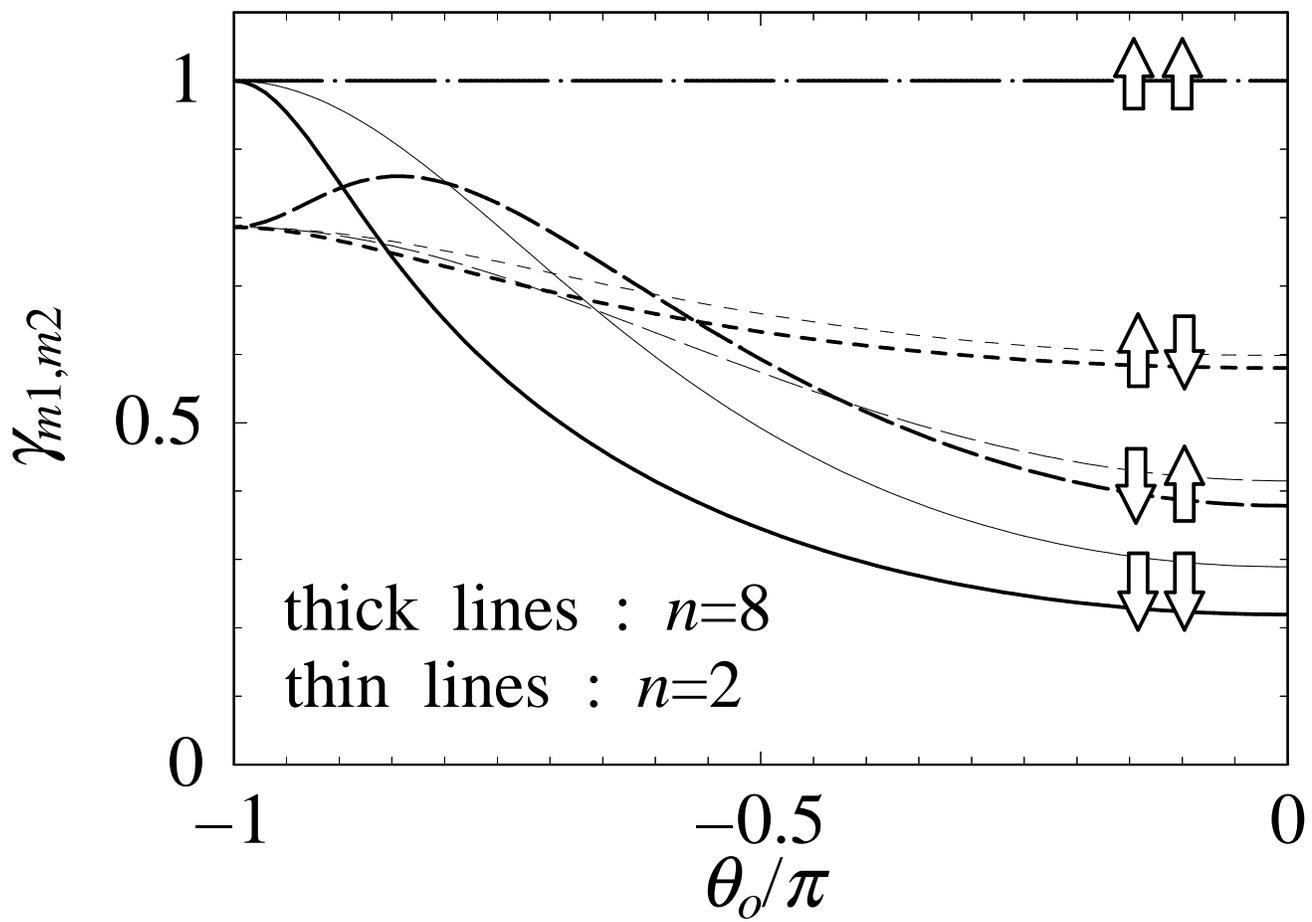
\end{document}